\documentclass{czjphys}

\def\POZ{\footnote{Presented at the $15^{\mathrm{th}}$ International
Colloquium on ``Integrable systems and Quantum symmetries",
Prague, 15--17 June 2006.}}

\newcommand{\hepth}[1]{{\tt hep-th/#1}}

\slacs{.4ex}

\begin{document}
\title{Purely transmitting integrable defects\,\POZ}
\authori{E. Corrigan}      \addressi{Department of Mathematics, University of York,
York YO10 5DD, UK}
\authorii{}     \addressii{}
%
\headauthor{E. Corrigan}            
\headtitle{Purely transmitting integrable defects}             
\lastevenhead{E. Corrigan: Purely transmitting integrable defects} 
\pacs{02.30Ik, 05.45Yv}     
\keywords{integrable field theory, solitons, defects} 
\refnum{A}
\daterec{11 July 2006}    
\issuenumber{0}  \year{2006} \setcounter{page}{1}
\maketitle

\begin{abstract}
Some aspects of integrable field theories possessing purely transmitting
defects are described. The main example is the sine-Gordon model and
several striking features of a classical field theory containing one
or more defects are pointed out. Similar features appearing in the
associated quantum field theory are also reviewed briefly.
\end{abstract}

\section{Classical picture}

The context of this topic is two-dimensional (i.e. one space --
one time) integrable classical (or quantum) field theory, and the
basic question concerns how to `sew' together field theories
defined in different segments of space.

\subsection{The setup}

The simplest situation has two scalar fields, $u(x,t)$, $x<x_0$
and $v(x,t)$, $x>x_0$, with a Lagrangian density given formally by
\begin{equation}
\label{lagrangian} {\cal L}=\theta(x_0-x){\cal L}_u +
\theta(x-x_0){\cal L}_v +\delta (x-x_0) {\cal B}(u,v)\,.
\end{equation}
The first two terms are the bulk Lagrangian densities for the
fields $u$ and $v$ respectively, while the third term provides the
sewing conditions; it could in principle  depend on $u$, $v$,
$u_t$, $v_t$, $u_x$, $v_x$, \ldots{} but the interesting question
is how to choose ${\cal B}$ so that the resulting system remains
integrable \cite{bczlandau}.

With free fields, there are many ways to choose ${\cal B}$. For example,
\begin{equation}
\label{deltaimpurity} {\cal B}(u,v)=-\sfrac12\,\sigma
uv+\sfrac{1}{2}(u_x+v_x)(u-v)\,,
\end{equation}
with standard choices for the bulk Lagrangians, leads to the
following set of field equations and sewing conditions,
\begin{equation}
\begin{array}{rcll}
(\partial^2+m^2)u&=&0\,,\quad&x<x_0\,,\\
(\partial^2+m^2)v&=&0\,,\quad&x>x_0\,,\\
u&=&v\,,\quad&x=x_0\,,\\
v_x-u_x&=&\sigma u\,,\quad& x=x_0\,,
\end{array}
\end{equation}
implying the fields are continuous with a discontinuity in the
derivative. This is an example of a $\delta$-impurity. Typically,
the sewing conditions at $x=x_0$ lead to reflection and
transmission and (for $\sigma <0$) a bound state. However, if the
fields on either side have nonlinear but integrable interactions
(e.g. each is a sine-Gordon field), the $\delta$-impurity destroys
the integrability (but still interesting \cite{Goodman02}).

If both $u$ and $v$ are sine-Gordon fields a suitable choice of
Lagrangian would be to take
\begin{equation}\label{D}
\begin{array}{rcl}
{\cal B}(u,v)&=&\frac{1}{2}(vu_t-uv_t)+{\cal D}(u,v)\,,\\[9pt]
{\cal D}(u,v)&=&\disty
-2\biggl(\sigma\cos\frac{u+v}2+\frac{1}{\sigma} \cos \frac{u-v}{2}\biggr)\\
\end{array}
\end{equation}
leading to the set of equations
\begin{equation}\label{sG}
\begin{array}{rcll}
\partial^2 u&=&-\sin u\,,\quad&x<x_0\,,\\
\partial^2 v&=&-\sin v\,,\quad&x>x_0\\[6pt]
u_x&=&\disty
v_t-\sigma \sin \frac{u+v}{2}-\frac{1}{\sigma}\sin\frac{u-v}{2}\,,\quad&x=x_0\\[9pt]
v_x&=&\disty
u_t+\sigma\sin\frac{u+v}{2}-\frac{1}{\sigma}\sin\frac{u-v}{2}\,,\quad&x=x_0\,.
\end{array}
\end{equation}
(The bulk coupling and mass parameter have been scaled away for
convenience.) This set up is not at all the same as the
$\delta$-impurity: it is integrable, there is no bound state for
any value of the parameter $\sigma$, and typically
$u(x_0,t)-v(x_0,t)\ne 0$, implying a discontinuity in the fields.
Clearly, the equations (\ref{sG}) describe a `defect'
(occasionally called a `jump-defect' to distinguish it from other
types). Note also that the sewing conditions are strongly
reminiscent of a B\"acklund transformation, and would be a
B\"acklund transformation if they were not `frozen' at $x=x_0$
(see, for example, \cite{Backlund}. That this setup is integrable
can be verified by constructing  Lax pairs using techniques
similar to those described in \cite{bcdr} for boundary situations.

Since the setup (\ref{sG}) is local, it is clear there may be many
defects, with parameters $\sigma_i$, at different locations $x_i$
along the $x$-axis.

\subsection{Energy and momentum}

Time translation invariance is not violated by the defect and
therefore there is a conserved energy, which includes a
contribution from the defect itself. On the other hand, space
translation is violated by a defect and therefore momentum might
not be expected to be conserved even allowing for a contribution
from the defect. It is worth investigating this in more detail in
terms of the quantity ${\cal D}(u,v)$ appearing in (\ref{D}).

The momentum carried by the fields on either side of the defect at
$x=x_0$ is given by
\begin{equation}\label{momentum}
P=\int_{-\infty}^{x_0}\D x\,u_xu_t+\int_{x_0}^\infty\D x\,v_xv_t
\end{equation}
and thus, using the defect conditions coming from (\ref{D}),
\begin{equation}\label{}
u_x=v_t-\frac{\partial{\cal D}}{\partial u}\,,\quad
v_x=u_t+\frac{\partial{\cal D}}{\partial v}\,,\quad\mathrm{at}\;\,
x=x_0\,,
\end{equation}
one finds
\begin{equation}\label{Pdot}
\dot P=\left[-v_t \frac{\partial{\cal D}}{\partial u} -
u_t \frac{\partial{\cal D}}{\partial v}-V(u)+W(v)
+\frac{1}{2}\left(\frac{\partial{\cal D}}{\partial u}\right)^2 -
\frac{1}{2}\left(\frac{\partial{\cal D}}{\partial v}\right)^2 \right]_{x_0}.
\end{equation}
In this expression the fields on either side of the defect have
been allowed to have (possibly different) potentials. Clearly,
(\ref{Pdot}) is not generally a total time-derivative of a
functional of the two fields. However, it will be provided, at
$x=x_0$,
\begin{equation}\label{Dequations}
\frac{\partial^2{\cal D}}{\partial u^2}=\frac{\partial^2{\cal
D}}{\partial v^2}\,, \qquad \frac{1}{2}\left(\frac{\partial{\cal
D}}{\partial u}\right)^2 - \frac{1}{2}\left(\frac{\partial{\cal
D}}{\partial v}\right)^2 = V(u)-W(v)\,.
\end{equation}
This set of conditions is satisfied by the sine-Gordon defect
function (\ref{sG}). However, there are other solutions too, for
example Liouville--Liouville, Liouville--massless free,
free--free. In fact, in many cases investigated so far, including
cases with several scalar fields \cite{bczlandau,bcztoda}, it
turns out that the requirements of integrability coincide with the
requirement that there be a modified conserved momentum.

\subsection{Classical scattering and solitons}

It is not difficult to check that the free-field limit of the sine-Gordon
setup, given by
\begin{equation}
{\cal D}(u,v)\rightarrow \frac{\sigma}{4}(u+v)^2+\frac{1}{4\sigma}(u-v)^2,
\end{equation}
leads to conditions describing a purely transmitting jump-defect
(i.e. no reflection). Given that fact, it is natural to ask what
might happen with solitons in the nonlinear sine-Gordon model (for
details concerning solitons, see for example \cite{Scott73}.

A soliton travelling in the positive $x$ direction (rapidity
$\theta$) is given by expressions
\begin{equation}
\E^{\I u/2}=\frac{1+\I E}{1-\I E}\,,\quad  x<x_0\,;\qquad \E^{\I
v/2}=\frac{1+\I zE}{1-\I zE}\,,\quad  x>x_0\,,
\end{equation}
where
\begin{equation}
E=\E^{ax+bt+c}\,,\quad a=\cosh\theta\,,\quad
b=-\sinh\theta\,,\quad \hbox{with $\E^c$ real}.
\end{equation}
The defect conditions (\ref{sG}) are satisfied provided ($\sigma
=\E^{-\eta}$)
\begin{equation}
z=\frac{\E^{-\theta}+\sigma}{\E^{-\theta}-\sigma}\equiv \coth
\left(\frac{\eta-\theta}{2}\right),
\end{equation}
and it is worth noting that $z^2$ would represent the delay
experienced by a soliton of rapidity $\theta$ passing another of
rapidity $\eta$. As it is, the quantity $z$ may change sign,
meaning, in fact, that a soliton can convert to an anti-soliton,
or vice-versa, besides being delayed, or even absorbed. In the
latter case, the defect gains a unit of topological charge in
addition to storing the energy and momentum of the soliton; in the
former, the defect gains (or loses) two units of topological
charge. Because the defect potential has period $4\pi$, all evenly
charged defects have identical energy--momentum, as do all oddly
charged defects. A fascinating possibility associated with this
type of defect (if it can be realized in practice) would be the
capacity to control solitons (see, for example \cite{cz}). Several
defects affect progressing solitons independently; several
solitons approaching a defect (inevitably possessing different
rapidities) are affected independently, with at most one of the
components being absorbed. Notice, too, that the situation is not
time-reversal invariant owing to the presence of explicit time
derivatives in eqs(\ref{sG}). Starting with an odd charged defect,
energy--momentum conservation would permit a single soliton to
emerge. However, classically, there is nothing to determine the
time at which the decay of the defect would occur. In that
situation, quantum mechanics should supply a probability for the
decay --- and indeed it does.

\section{Quantum picture}

\subsection{The transmission matrix}

Following the remarks made in the last section one expects two
types of transmission matrix, one of them, $^{\rm even}T$,
referring to even-labelled defects --- and this is expected to be
unitary, since these defects cannot decay --- and the other,
$^{\rm odd}T$, referring to odd-labelled defects. The latter is
not expected to be unitary, yet would be expected to be related
(via a bootstrap principle) to a complex bound state pole in the
former. In fact this is precisely what happens and, remarkably
enough, the relevant transmission matrices were described by Konik
and LeClair some time ago \cite{Konik97}. Using roman labels to
denote soliton states (taking the value $\pm 1$), and greek labels
to label the charge on a defect, and assuming topological charge
is conserved in every process, it is expected that both
transmission matrices will satisfy `triangle' compatibility
relations with the bulk $S$-matrix, for example:
\begin{equation}
\label{STT} S_{ab}^{cd}(\theta_1-\theta_2)\,
T_{d\alpha}^{f\beta}(\theta_1)\,T_{c\beta}^{e\gamma}(\theta_2)=
T_{b\alpha}^{d\beta}(\theta_2)\,T_{a\beta}^{c\gamma}(\theta_1)\,
S_{cd}^{ef}(\theta_1-\theta_2)\,.
\end{equation}
Here, it is supposed the solitons are travelling along the
positive $x$-axis ($\theta_1>\theta_2>0$). The bulk $S$-matrix
depends on the bulk coupling $\beta$ via the quantity
$\gamma=8\pi/\beta^2 -1$, and the conventions used are those
adopted in \cite{bczsg}. The equations (\ref{STT}) are well known
in many contexts involving the notion of integrability (see
\cite{Jimbo}), but were discussed first with reference to  defects
by Delfino, Mussardo and Simonetti \cite{Delf94a}; if the
possibility of reflection were to be allowed an alternative
framework (such as the one developed by Mintchev, Ragoucy and
Sorba \cite{Mintchev02}), might be more appropriate. Here, the
defect is expected to be purely transmitting.

The solution (for general $\beta$, and for even or odd labelled
defects --- note the labelling is never mixed by (\ref{STT})), is
given by
\begin{equation}
\label{KL}
{\slacs{1.2ex}
T_{a\alpha}^{b\beta}(\theta)=f(q,x)\left(\begin{array}{cc}
\nu^{-1/2}Q^\alpha\delta_\alpha^\beta &
q^{-1/2}\E^{\gamma(\theta-\eta)}\delta_\alpha^{\beta-2}\\[5pt]
q^{-1/2}\E^{\gamma(\theta-\eta)}\delta_\alpha^{\beta+2}&
\nu^{1/2}Q^{-\alpha}\delta_\alpha^\beta\\
\end{array}\right).}
\end{equation}
In (\ref{KL}) a block form has been adopted with the labels $a$,
$b$ labelling the four block elements on the right hand side, and
where $\nu$ is a free parameter, as is $\eta$ (to be identified
with the defect parameter introduced in the previous section), and
\begin{equation}
q=\E^{\I\pi\gamma}\,,\quad x=\E^{\gamma\theta}\,,\quad
Q^2=-q=\E^{4\pi^2\I/\beta^2}\,.
\end{equation}
In addition, $^{\rm even}T$ is a unitary matrix (for real
$\theta$), and both types of transmission matrix must be
compatible with soliton--anti-soliton annihilation as a virtual
process. These two requirements place the following restrictions
on the overall factor for the even transmission matrix,
$^{\E}f(q,x)$:
\begin{equation}
\left\{\begin{array}{l}
{}^{\E}\bar f(q,x)={^{\E}f}(q,qx)\,,\\[5pt]
{}^{\E}f(q,x)\;{^{\E}\!f}(q,qx)\left(1+\E^{2\gamma(\theta-\eta)}\right)=1\,.
\end{array}\right.
\end{equation}
These do not determine $^{\E}f(q,x)$ uniquely but the `minimal'
solution determined by Konik--LeClair has
\begin{equation}
\label{KLf} {}^{\E}f(q,x)=
\frac{\E^{\I\pi(1+\gamma)/4}}{1+\I\E^{\gamma(\theta-\eta)}}\,
\frac{r(x)}{\bar r(x)}\,,
\end{equation}
with ($z=\I\gamma(\theta-\eta)/2\pi$),
\begin{equation}
r(x)=\prod_{k=0}^\infty\,
\frac{\Gamma(k\gamma+\sfrac14-z)\,\Gamma((k+1)\gamma+\sfrac34-z)}
{\Gamma((k+\sfrac12)\gamma+\sfrac14-z)\,\Gamma((k+\sfrac12)\gamma+\sfrac34-z)}\,.
\end{equation}
It is worth noting that the apparent pole in (\ref{KLf}) at
$1+\I\E^{\gamma(\theta-\eta)}=0$ is actually
cancelled by a pole at the same location in $\bar r(x)$. However, there is another pole
at
\begin{equation}
\theta=\eta -\frac{\I\pi}{2\gamma} \rightarrow \eta\;\;{\rm
as}\;\;\beta\rightarrow 0\,,
\end{equation}
uncancelled by a zero, and this does actually represent the
expected unstable bound state alluded to in the first section.

Several brief remarks are in  order. It is clear, on examining
(\ref{KL}), that the processes in which a classical soliton would
inevitably convert to an anti-soliton are clearly dominant even in
the quantum theory, yet suppressed if a classical soliton is
merely delayed. This much is guaranteed by the factor
$\E^{\gamma(\theta-\eta)}$ appearing in the off-diagonal terms. A
curious feature is the different way solitons and anti-solitons
are treated by the diagonal terms in (\ref{KL}). They are treated
identically by the bulk $S$-matrix yet one should not be surprised
by this since the classical defect conditions (\ref{sG}) do not
respect all the usual discrete symmetries. Indeed, the dependence
of the diagonal entries on the bulk coupling can be demonstrated
to follow from  the classical picture by using a functional
integral type of argument, as explained more fully in
\cite{bczsg}. The sine-Gordon spectrum contains bound states
(breathers), and it is interesting to calculate their transmission
factors. This much has been done \cite{bczsg}. However, it would
also be interesting to attempt to match these breather
transmission factors to perturbative calculations, and this has
not yet been done. There are also open questions concerning how to
treat defects in motion. From a classical perspective it seems
quite natural that defects might move and scatter \cite{bczsg},
however it is less clear how to describe this in the quantum field
theory, or indeed to understand what these objects really are.

It is quite remarkable that the simple-looking question asked at
the beginning has led to an interesting avenue of enquiry that
does not appear to have been explored previously, that links with
results, such as (\ref{KL}), which had been obtained for seemingly
quite different reasons, and that is not yet exhausted (for
example, see \cite{Gomes}, for an extension to supersymmetric
sine-Gordon).
\bigskip

\noindent{\small\textbf{Acknowledgements.} I am very grateful to
the organisers for giving me the opportunity to review this
material, to Peter Bowcock and to Cristina Zambon for many
discussions concerning this topic and for a longstanding
collaboration, and to several other members of EUCLID, a Research
Training Network funded by the European Commission (contract
number HPRN-CT-2002-00325).}

\end {document}